\documentclass[a4paper,10pt]{article}

\usepackage{graphicx}
\usepackage{epsfig}
\usepackage{amssymb,amsmath}
\usepackage{pstricks}
\usepackage[utf8]{inputenc}
\usepackage{cite}

\input{epsf.sty}

\begin{document}

\begin{titlepage}

\begin{flushright}
BI-TP-10/04\\
MS-TP-10-01
\end{flushright}
\begin{centering}
\vfill

{\bf\Large The pressure of strong coupling lattice QCD with heavy quarks, 
the hadron resonance gas and the large N limit}

\vspace{0.8cm}
 
Jens Langelage$^1$ and Owe Philipsen$^2$

\vspace{0.3cm}
{\em 
$^1$ Fakult\"at f\"ur Physik, Universit\"at Bielefeld, \\
33615 Bielefeld, Germany}

\vspace{0.3cm}
{\em $^2$
Institut f\"ur Theoretische Physik, Westf\"alische Wilhelms-Universit\"at M\"unster,\\
48149 M\"unster, Germany}
\vspace*{0.7cm}
 
\begin{abstract}
In this paper we calculate the pressure of pure lattice Yang-Mills theories and lattice QCD with heavy quarks by means of strong coupling expansions. Dynamical fermions are introduced with a hopping parameter expansion, which also allows for the incorporation of finite quark chemical potential.
We show that in leading orders the results are in full agreement with expectations from the hadron resonance gas model, thus validating it with a first principles calculation. For pure Yang-Mills theories we obtain the corresponding ideal glueball gas, in QCD with heavy quarks our result equals that of an ideal gas of mesons and baryons.
Another finding is that the Yang-Mills pressure in the large N limit is of order $\sim N^0$ to the calculated orders, when the inverse 't Hooft coupling is used as expansion parameter. This property is expected in the confined phase, where our calculations take place.
\end{abstract}
\end{centering}

\noindent
\vfill
\noindent

\end{titlepage}

\section{Introduction}

With the experimental heavy ion programmes at RHIC and LHC, it is possible to 
perform experimental research on the properties of the quark gluon plasma. 
In order to distinguish this new matter phase from ordinary matter, 
it is decisive to have an accurate knowledge of both. 
While in the deconfined phase qualitative insights are often based on perturbation
theory,
one of the main theoretical frameworks used to describe the hadronic phase, besides e.g. chiral perturbation theory, is the hadron 
resonance gas model and its variants. 
The idea is to model the confined phase as an ideal gas of hadrons and resonances. 
Confronting predictions based on this picture 
with non-perturbative results from lattice simulations shows good agreement up to 
the phase transition region \cite{krt1,krt2,meyer,huov}.

At present, the hadron resonance gas is a model requiring a lot of experimental
input data on hadronic properties in order to give predictions, 
and it lacks a proper field theoretic motivation.
In this paper we derive the hadron resonance gas model 
as an appropriate effective theory for strong gauge couplings 
from the QCD Lagrangian with quarks and gluons as the fundamental degrees of freedom.
In particular, we calculate the pressure 
- or the negative free energy density - in a combined strong coupling and 
hopping parameter expansion in the lattice theory.  
These expansions are known to give convergent series. Strong coupling 
methods have also been used, e.g., to employ the properties of thermal QCD with 
mean field methods \cite{ohnishi1,ohnishi2,ohnishi3}, 
numerical simulations \cite{forc} or series expansion methods \cite{Jako,lang,La}.

The outline of this paper is the following: after a short review of the 
hadron resonance gas model and its description on the lattice, we calculate 
the pressure for pure $SU(N)$ lattice Yang-Mills theories, 
with special emphasis on the cases $N=3$ and $N\rightarrow\infty$. 
Then we introduce dynamical quarks via a hopping parameter expansion 
and calculate the pressure for QCD with heavy quarks. In all cases we obtain the 
right spin degeneracies for the lowest mass hadrons and the correct 
dependence on chemical potential when compared to predictions of 
the hadron resonance gas model.

\section{Hadron resonance gas}

The basic quantity of thermal systems is the grand partition function \cite{kapusta}
\begin{eqnarray}
Z=\mathrm{tr}\,\mathrm{e}^{-(H-\mu N)T^{-1}},
\end{eqnarray}
from which we can deduce the pressure as
\begin{eqnarray}
P=\frac{\partial(T\ln\,Z)}{\partial V}\qquad\mathop{\longrightarrow}_{V\rightarrow\infty}\qquad P=\frac{T}{V}\ln\,Z.
\label{eq_press}
\end{eqnarray}
The last equality holds in the infinite volume limit, which we will always consider in the following.

In the ideal gas picture, we can write the partition function as a sum over one-particle partition functions of the hadrons and resonances \cite{krt1,krt2}
\begin{eqnarray}
\ln\,Z=\sum_i\ln\,Z_i^1.
\label{eq_1p}
\end{eqnarray}
In the continuum, the one-particle partition functions at temperature $T$ 
and chemical potential $\mu_i$ are given by
\begin{eqnarray}
\ln\,Z_i^1=Vg_i\mathop{\int}_{-\infty}^{\infty}\mathop{\int}_{-\infty}^{\infty}\mathop{\int}_{-\infty}^{\infty}\frac{d^3p}{(2\pi)^3}\eta\ln\,\left(1+\eta\,\mathrm{e}^{-(\omega_i-\mu_i)T^{-1}}\right),
\end{eqnarray}
where $g_i$ is a possible spin-isospin degneracy factor and $\omega_i=\sqrt{p^2+m_i^2}$ for particles of mass $m_i$. The factor $\eta$ refers to bosons ($\eta=-1$) or fermions ($\eta=1$). With Eqs~(\ref{eq_press}) and (\ref{eq_1p}) the pressure of the hadron resonance gas is given as a sum over one-particle contributions
\begin{eqnarray}
P=\sum_iP_i.
\end{eqnarray}
Calculating the ideal gas pressure on the lattice, we have to keep in mind that 
the momentum integration is limited to the first Brillouin zone, i.e.~to the 
interval $\big[-\frac{\pi}{a},\frac{\pi}{a}\big)$. Hence the pressure is given by 
\begin{eqnarray}
P_i=\frac{g_i}{L_{\tau}}\mathop{\int}_{-\frac{\pi}{a}} ^\frac{\pi}{a}\mathop{\int}_{-\frac{\pi}{a}} ^\frac{\pi}{a}\mathop{\int}_{-\frac{\pi}{a}} ^\frac{\pi}{a}\frac{d^3p}{(2\pi)^3}\eta\ln\left[1+\eta\mathrm{e}^{-\left(\omega_i-\mu_i\right)L_{\tau}}\right],
\end{eqnarray}
where the temperature is related to the temporal lattice extent via
\begin{eqnarray}
T=\frac{1}{L_{\tau}}=\frac{1}{N_{\tau}a}.
\end{eqnarray}
For heavy particles it is convenient to use non-relativistic approximations. 
In the static limit we have $\omega\simeq m$.
However, in the continuum this approximation leads to
a divergent momentum integration. Moreover, we have to 
integrate over arbitrarily large momenta where the approximation does not hold. 
Thus we have to expand the relativistic dispersion relation at least 
to $\omega\simeq m+\frac{p^2}{2m}$ in order to render the integral finite.

Using the static approximation on the lattice, 
the momentum integration gets trivial and we obtain
\begin{eqnarray}
P_i=\frac{g_i}{L_{\tau}a^3}
\eta\ln\left[1+\eta\mathrm{e}^{-\left(m_i-\mu_i\right)L_{\tau}}\right].
\end{eqnarray}
In the following we will use the hopping parameter expansion, which is applicable 
for heavy quarks and thus for heavy composite particles like the 
corresponding mesons and baryons. Hence it is justified to consider only the limit
\begin{eqnarray}
m\gg T=\frac{1}{L_{\tau}},
\end{eqnarray}
which yields
\begin{eqnarray}
P=\sum_iP_i=\frac{1}{L_{\tau}a^3}\sum_ig_i\mathrm{e}^{-(m_i-\mu_i)L_{\tau}}.
\end{eqnarray}
Using the fact that to each baryon with $\mu_B$ there is an antibaryon with $-\mu_B$, we obtain
\begin{eqnarray}
P=\frac{1}{L_{\tau}a^3}\sum_Mg_M\mathrm{e}^{-m_ML_{\tau}}+\frac{1}{L_{\tau}a^3}\sum_Bg_B\mathrm{e}^{-m_BL_{\tau}}\cosh(\mu_BL_{\tau}),
\label{eq_hrglatt}
\end{eqnarray}
where we have split the sum into a mesonic and a baryonic part.
It is this form of the pressure which we will derive in the following from a first principles strong coupling calculation.

\section{Pressure of lattice Yang-Mills theories}

In an earlier paper \cite{La} the general procedure to calculate the pressure 
in case of $SU(2)$ has already been discussed. Thus we will only shortly 
outline the derivation of the $SU(N\geq3)$ results and focus 
on the more interesting cases $N=3$ and the limit $N\rightarrow\infty$.

To obtain the physical pressure, we need to calculate
\begin{eqnarray}
P_{\mathrm{phys}}=P(L_{\tau})-P(\infty),
\end{eqnarray}
where the subtraction of vacuum contributions ($L_\tau\rightarrow \infty$) 
is necessary for renormalisation. On a lattice with temporal extent $L_\tau$ the
expression to evaluate is
\begin{eqnarray}
P(L_{\tau})=\frac{6}{a^4}\ln\,c_0(\beta)+\frac{1}{L_{\tau}V}\sum_{C=\left(X_i^{n_i}\right)}a(C)\prod_i\Phi(X_i)^{n_i}.
\end{eqnarray} 
Here $\Phi(X_i)$ is the contribution of a graph $X_i$, $C$ specifies a cluster of
graphs and $a(C)$ is a combinatorial factor determined by the formalism of moments 
and cumulants \cite{mm}.
When $L_\tau$ goes to infinity this is the zero temperature contribution.
\begin{figure}
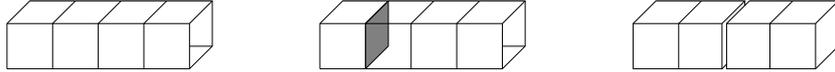

\begin{center}
\begin{minipage}{4cm}
\scalebox{0.3}{
\psline(0,0)(8,0)(9,1)(9,3)(1,3)(0,2)(0,0)
\psline(0,2)(8,2)(9,3)
\psline(2,0)(2,2)(3,3)
\psline(4,0)(4,2)(5,3)
\psline(6,0)(6,2)(7,3)
\psline(8,0)(8,2)
\psline(9,1)(8,1)
}
\end{minipage}
\begin{minipage}{4cm}
\scalebox{0.3}{
\psline[fillcolor=gray,fillstyle=solid](2,0)(3,1)(3,3)(2,2)(2,0)
\psline(0,0)(8,0)(9,1)(9,3)(1,3)(0,2)(0,0)
\psline(0,2)(8,2)(9,3)
\psline(2,0)(2,2)(3,3)
\psline(4,0)(4,2)(5,3)
\psline(6,0)(6,2)(7,3)
\psline(8,0)(8,2)
\psline(9,1)(8,1)
}
\end{minipage}
\begin{minipage}{4cm}
\scalebox{0.3}{
\psline(0,0)(3.9,0)(3.9,2)(4.9,3)(1,3)(0,2)(0,0)
\psline(2,0)(2,2)(3,3)
\psline(0,2)(3.9,2)
\psline(4.9,3)(4.9,2.8)
\psline(3.9,0)(4.1,0.2)
\psline(4.1,0)(8,0)(9,1)(9,3)(5.1,3)(4.1,2)(4.1,0)
\psline(6,0)(6,2)(7,3)
\psline(8,0)(8,2)(9,3)
\psline(4.1,2)(8,2)
}
\end{minipage}
\caption{Graphs that appear on the different lattices. Left: Leading order tube for $N_{\tau}=4$. 
Middle: Graph with an inner plaquette in the case $N_{\tau}=4$. 
Right: Graph with two slits contributing on both, $N_{\tau}=4,\infty$, lattices.}\label{tubes}
\end{center}
\end{figure}
The leading order contribution to the subtracted pressure is a tube of $4N_{\tau}$ fundamental plaquettes spanning around the temporal direction, as shown in Fig.~\ref{tubes} (left). In contrast to $SU(2)$ we have two fundamental representations in the character expansion and counting the 3 different cross sections of the tube per site we get
\begin{eqnarray}
P=\frac{6}{L_{\tau}a^3}u^{4N_{\tau}}\left(1+{\cal{O}}\left(u^2\right)\right),
\label{eq_lo}
\end{eqnarray}
where $u$ is the character of the fundamental representation and effective expansion
parameter. A volume factor $N_s^3$ cancels out, 
due to the spatial translations of the tube. 
This is the leading order result for all $N\geq3$. An explicit 
dependence on the gauge group appears in higher orders, but is indirectly 
encoded in the functional behaviour
\begin{eqnarray}
u(\beta)=\frac{\beta}{2N^2}+\ldots.
\end{eqnarray}
It is interesting to note that for $N\rightarrow\infty$ we have the correspondence \cite{drouffe}
\begin{eqnarray}
u(\beta)\equiv\frac{\beta}{2N^2}=\frac{1}{g^2N}=\lambda^{-1},
\end{eqnarray}
and hence the expansion parameter of the fundamental representation equals the inverse of the 't Hooft coupling $\lambda$ \cite{thooft}.

\begin{figure}
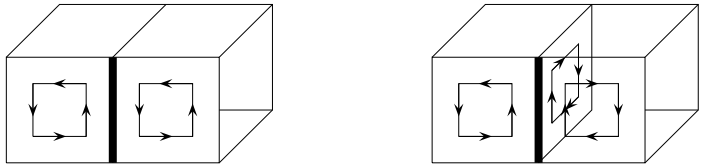

\hspace{3cm}
\scalebox{0.35}{
\psline(0,0)(8,0)(8,4)(0,4)(0,0)
\psline(8,0)(10,2)(10,6)(2,6)(0,4)
\psline(4,0)(4,4)(6,6)
\psline(8,4)(10,6)
\psline(10,2)(8,2)
\psline(1,1)(3,1)(3,3)(1,3)(1,1)
\psline(5,1)(7,1)(7,3)(5,3)(5,1)
\psline[arrowsize=10pt]{->}(1,1)(2.25,1)
\psline[arrowsize=10pt]{->}(3,1)(3,2.25)
\psline[arrowsize=10pt]{->}(3,3)(1.75,3)
\psline[arrowsize=10pt]{->}(1,3)(1,1.75)
\psline[arrowsize=10pt]{->}(5,1)(6.25,1)
\psline[arrowsize=10pt]{->}(7,1)(7,2.25)
\psline[arrowsize=10pt]{->}(7,3)(5.75,3)
\psline[arrowsize=10pt]{->}(5,3)(5,1.75)
\psline[linewidth=.3](4,0)(4,4)
\psline(16,0)(24,0)(24,4)(16,4)(16,0)
\psline(24,0)(26,2)(26,6)(18,6)(16,4)
\psline(20,0)(20,4)(22,6)
\psline(24,4)(26,6)
\psline(26,2)(24,2)
\psline(20,0)(22,2)(22,6)
\psline(17,1)(19,1)(19,3)(17,3)(17,1)
\psline(21,1)(23,1)(23,3)(21,3)(21,1)
\psline(20.5,1.5)(20.5,3.5)(21.5,4.5)(21.5,2.5)(20.5,1.5)
\psline[arrowsize=10pt]{->}(17,1)(18.25,1)
\psline[arrowsize=10pt]{->}(19,1)(19,2.25)
\psline[arrowsize=10pt]{->}(19,3)(17.75,3)
\psline[arrowsize=10pt]{->}(17,3)(17,1.75)
\psline[arrowsize=10pt]{->}(21,1)(21,2.25)
\psline[arrowsize=10pt]{->}(21,3)(22.25,3)
\psline[arrowsize=10pt]{->}(23,3)(23,1.75)
\psline[arrowsize=10pt]{->}(23,1)(21.75,1)
\psline[arrowsize=10pt]{->}(20.5,1.5)(20.5,2.75)
\psline[arrowsize=10pt]{->}(20.5,3.5)(21,4)
\psline[arrowsize=10pt]{->}(21.5,4.5)(21.5,3.25)
\psline[arrowsize=10pt]{->}(21.5,2.5)(21,2)
\psline[linewidth=.3](20,0)(20,4)
}
\caption{Orientation of plaquettes that meet at some link (thick line) in case of $SU(3)$. Left: Fundamental and antifundamental representation at the link resulting in a conserved orientation of the corresponding plaquettes. Right: Three fundamental representations at the link and a change of orientation, which has to be cancelled to give an allowed graph.}\label{fig_ori}
\end{figure}

The leading corrections to the basic polymer arise from putting spacelike plaquettes in higher dimensional representations inside the torus, as in Fig.~\ref{tubes} (middle). In order to have non-vanishing contribution according to the integration formula
\begin{eqnarray} 
\int dU \chi_r(U)=\delta_{r,0},
\label{trivial}
\end{eqnarray}
the allowed representations are those whose conjugates appear either in $f\otimes\overline{f}$ or in $f\otimes f$, the Kronecker products of the fundamental representations. The latter representations are forced to appear an even number of times along the tube, since they induce a change of orientation of the outer plaquettes and this has to be cancelled due to the periodic boundary conditions, cf. Fig.~\ref{fig_ori}. If orientation changing plaquettes appear $n$ times, we take this fact into account by inserting a factor
\begin{eqnarray}
\frac{1}{2}\left[1+(-1)^n\right]\label{eq_ff} 
\end{eqnarray}
into the summation of these contributions. Let us denote the parameters of the 
representations appearing in $f\otimes f$ with $v_1$ and $v_2$ and their respective
dimensions $d_{v_1},d_{v_2}$. In general,
\begin{eqnarray}
d_{v_1}v_1&=&\frac{N(N+1)}{2}\,v_1=\frac{1}{2}\left(Nu\right)^2\label{eq3}+\ldots\\\nonumber
\\
d_{v_2}v_2&=&\frac{N(N-1)}{2}\,v_2=\frac{1}{2}\left(Nu\right)^2+\ldots.\label{eq_v}
\end{eqnarray}
In the case of $SU(3)$ we have to identify $v_2$ with the complex conjugate fundamental 
representation ($3\otimes3=\overline{3}\oplus6$), hence in this case $v_2$ equals 
$u$ rather than being of ${\cal{O}}(u^2)$.
The parameter of the adjoint representation appearing in $f\otimes\overline{f}$, which we will call $w$, fulfills
\begin{eqnarray}
d_{ad}w&=&(N^2-1)\,w=\left(Nu\right)^2+\ldots\label{eq_w}.
\end{eqnarray}
Another correction is to add slits to the tube 
by inserting two fundamental plaquettes at the same location, 
c.f.~Fig.~\ref{tubes} (right). For finite $L_\tau$ 
the minimal number of slits is two, since no plaquette is allowed to be
occupied twice by the same polymer. This insertion gives an additional factor of $2N^2u^2$. The factor of two appears because adding a slit increases the number of polymers by one and this new polymer can contribute in two different orientations. Thus we get a chain of polymers with a combinatorial factor $a(C)$ of the resulting cluster $C$, which depends on whether the chain polymer is embedded in the finite lattice with periodic boundary conditions or in the infinite lattice. It is given by
\begin{eqnarray}
\mbox{Finite lattice, pbc:}&&a(C)=(i-1)(-1)^{i+1}\nonumber\\
\mbox{Infinite lattice:}&&a(C)=(-1)^{i},
\end{eqnarray}
where $i$ counts the number of polymers.
Now we sum over all possible contributions of the torus with inner decorations. 
This calculation proceeds in the same way as in the case of $SU(2)$ \cite{La}, 
with the difference of inserting the factor Eq.~(\ref{eq_ff}).
The result splits into two terms,
\begin{eqnarray}
P&=&\frac{3}{N_{\tau}}u^{4N_{\tau}}\left[c_N^{N_{\tau}}+b_N^{N_{\tau}}\right]\left(1+{\cal{O}}\left(u^4\right)\right).
\end{eqnarray}
We have introduced
\begin{eqnarray}
c_N&=&1+d_{v_1}v_1+d_{v_2}v_2+d_{ad}w-2d_f^2u^{2}
\\
b_N&=&1-d_{v_1}v_1-d_{v_2}v_2+d_{ad}w
\end{eqnarray}
as gauge group dependent factors.
In the limit $N\rightarrow\infty$ the expansion parameters simplify \cite{drouffe}
\begin{eqnarray}
d_{v_1}v_1&=&\frac{1}{2}(Nu)^2,\nonumber\\
d_{v_2}v_2&=&\frac{1}{2}(Nu)^2,\nonumber\\
d_{ad}w&=&(Nu)^2,
\end{eqnarray}
and these gauge group factors become trivial
\begin{eqnarray}
c_{\infty}&=&1+\frac{1}{2}(Nu)^2+\frac{1}{2}(Nu)^2+(Nu)^2-2N^2u^2=1\nonumber\\
b_{\infty}&=&1-\frac{1}{2}(Nu)^2-\frac{1}{2}(Nu)^2+(Nu)^2=1.
\end{eqnarray}
As a consequence the contributions of inner plaquettes cancel against each other. 
This gives a first hint that in the confined phase the pressure is a 
quantity of ${\cal{O}}(N^0)$, and thus finite in the limit $N\rightarrow\infty$, 
when expressed in terms of the inverse 't Hooft coupling $u$.
\begin{figure}
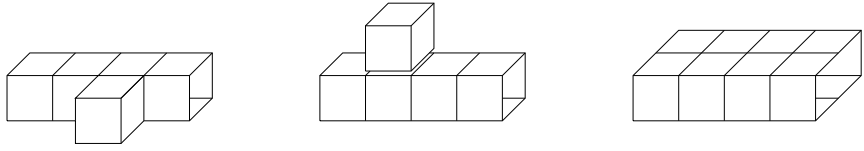

\begin{center}
\begin{minipage}{4cm}
\scalebox{0.3}{
\psline(0,0)(3,0)(3,-1)(5,-1)(6,0)(8,0)(9,1)(9,3)(1,3)(0,2)(0,0)
\psline(0,2)(4,2)(3,1)(5,1)(6,2)(8,2)
\psline(2,0)(2,2)(3,3)
\psline(5,3)(4,2)(6,2)
\psline(5,-1)(5,1)(7,3)
\psline(6,0)(6,2)
\psline(8,0)(8,2)(9,3)
\psline(3,0)(3,1)
\psline(9,1)(8,1)
}
\end{minipage}
\begin{minipage}{4cm}
\scalebox{0.3}{
\psline(0,0)(8,0)(9,1)(9,3)(5,3)(4,2)(0,2)(0,0)
\psline(2,2.2)(4,2.2)(5,3.2)(5,5.2)(3,5.2)(2,4.2)(2,2.2)
\psline(4,2.2)(4,4.2)(5,5.2)
\psline(2,4.2)(4,4.2)
\psline(0,2)(1,3)(2,3)
\psline(2,0)(2,2)(2.2,2.2)
\psline(4,0)(4,2)(8,2)(9,3)
\psline(6,0)(6,2)(7,3)
\psline(8,0)(8,2)
\psline(5,3)(4.8,3)
\psline(9,1)(8,1)
}
\end{minipage}
\begin{minipage}{4cm}
\scalebox{0.3}{
\psline(0,0)(8,0)(10,2)(10,4)(2,4)(0,2)(0,0)
\psline(2,0)(2,2)(4,4)
\psline(4,0)(4,2)(6,4)
\psline(6,0)(6,2)(8,4)
\psline(8,0)(8,2)(10,4)
\psline(0,2)(8,2)
\psline(1,3)(9,3)
\psline(9,1)(8,1)
\psline(10,2)(8,2)
}
\end{minipage}
\caption{Correction graphs. Left: One shifted plaquette. Middle: One added cube. 
Right: Graph with a perimeter of six. To all graphs one can also add inner 
plaquettes and slits.}\label{graphs}
\end{center}
\end{figure}

It is in complete analogy to $SU(2)$ to calculate higher order corrections. 
One can decorate the leading polymer, either geometrically or using higher 
dimensional plaquettes, one can add whole new polymers or one can combine these 
possibilities. Examples are shown in Fig.~\ref{graphs}. 
Again we have contributions of some polymers only for large $N_{\tau}$, e.g.~the 
graph in Fig.~\ref{graphs} (left), which is absent for $N_\tau=1$. Some polymers only 
contribute for small $N_{\tau}$, since their correction is of order $u^{2N_{\tau}}$ 
or higher, like the graph in Fig.~\ref{graphs} (right). 
The results for the series up to ${\cal{O}}(u^8)$ in the correction 
for the cases $N=3$ and $N\rightarrow\infty$ are given in the appendix.

\subsection{Hadron resonance gas for $SU(3)$}

In the case of $SU(3)$ the three lowest lying glueball states, which can be 
extracted from the strong coupling series of 
plaquette correlators, have masses of \cite{mu,seo}
\begin{eqnarray}
am_1(A)&=&-4\ln\,u-\ln\left(1+3u+6v_1+8w-18u^2\right)+{\cal{O}}(u^4)\\
am_2(E)&=&-4\ln\,u-\ln\left(1+3u+6v_1+8w-18u^2\right)+{\cal{O}}(u^4)\\
am_3(T)&=&-4\ln\,u-\ln\left(1-3u-6v_1+8w\right)+{\cal{O}}(u^4).
\end{eqnarray}
The subscript denotes the spin degeneracy of the glueball. 
Rewriting our result of the pressure
\begin{eqnarray}
P=\frac{3}{L_{\tau}a^3}u^{4N_{\tau}}\left[c_3^{N_{\tau}}+b_3^{N_{\tau}}\right]\left(1+{\cal{O}}\left(u^4\right)\right),
\end{eqnarray}
we obtain
\begin{eqnarray}
P=\frac{1}{L_{\tau}a^3}\left[\mathrm{e}^{-m_1(A)L_{\tau}}+2\mathrm{e}^{-m_2(E)L_{\tau}}+3\mathrm{e}^{-m_3(T)L_{\tau}}\right]\left(1+{\cal{O}}\left(u^4\right)\right),
\end{eqnarray}
which has to be compared with Eq.~(\ref{eq_hrglatt}). 
We see that the hadron resonance gas result equals the one derived with 
strong coupling methods to the orders considered here. Corrections to the ideal hadron gas results appear in ${\cal{O}}(u^4)$.

The strong coupling expansion of the pressure can also be used for the determination of the lower integration constant in the so-called integral method. In numerical simulations a direct computation of the partition function is not possible \cite{Boyd}, so that one has to calculate derivatives to get physical quantities. The pressure can be calculated, e.g. as an $\beta$-integral
\begin{eqnarray}
\frac{P}{T^4}\Bigg\vert_{\beta_0}^{\beta}=N_\tau^4\mathop{\int}_{\beta_0}^{\beta}d\beta^\prime\left[6{\cal{P}}_0-3({\cal{P}}_\tau+{\cal{P}}_s)\right],
\end{eqnarray}
where ${\cal{P}}_{s,\tau}$ are the expectation values of space-space and space-time plaquettes on an asymmetric lattice $N_\tau<N_s$ and ${\cal{P}}_0$ is the one of a symmetric lattice $N_\tau=N_s$. The lower integration constant $P(\beta_0)$ has to be fixed by hand. Usually one argues with an exponentially small pressure $P\sim\mathrm{exp}(-m_G/T)$ below $T_c$, where $m_G$ is the lowest glueball mass, and chooses $\beta_0$ correspondingly. With our calculations we have justified this assumption, but it is also possible to use our series expansion to consider some specific value $P(\beta_0)$ for larger values of $\beta_0$.

\subsection{The limit $N\rightarrow\infty$}

\begin{center}
\begin{figure}[t!]
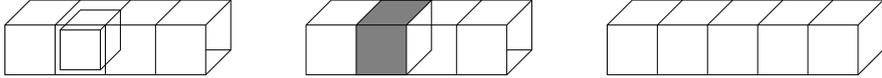

\hspace{2cm}
\scalebox{0.33}{
\psline(0,0)(8,0)(9,1)(9,3)(1,3)(0,2)(0,0)
\psline(2,0)(2,2)(3,3)
\psline(4,0)(4,2)(5,3)
\psline(6,0)(6,2)(7,3)
\psline(0,2)(8,2)(9,3)
\psline(8,0)(8,2)
\psline(9,1)(8,1)
\psline(2.2,0.2)(3.8,0.2)(3.8,1.8)(2.2,1.8)(2.2,0.2)
\psline(3.8,0.2)(4.6,1)(4.6,2.6)(3,2.6)(2.2,1.8)
\psline(3.8,1.8)(4,2)
\psline(12,0)(20,0)(21,1)(21,3)(13,3)(12,2)(12,0)
\psline(18,0)(18,2)(19,3)
\psline(20,0)(20,2)(21,3)
\psline(12,2)(14,2)
\psline(16,2)(20,2)
\psline(17,3)(17,1)(16,0)
\psline(21,1)(20,1)
\pspolygon[fillstyle=solid,fillcolor=gray](14,0)(14,2)(15,3)(17,3)(16,2)(16,0)(14,0)
\psline(14,2)(16,2)
\psline(24,0)(34,0)(35,1)(35,3)(25,3)(24,2)(24,0)
\psline(26,0)(26,2)(27,3)
\psline(28,0)(28,2)(29,3)
\psline(30,0)(30,2)(31,3)
\psline(32,0)(32,2)(33,3)
\psline(24,2)(34,2)(35,3)
\psline(34,0)(34,2)
}
\caption{Polymers contributing to ${\cal{O}}\left(u^6\right)$ in the correction. 
Left: One cube has been added inside the tube. 
Middle: Four fundamental plaquettes have been changed for plaquettes $\sim u^2$. 
Two fundamental plaquettes had to be added at the inside. 
Right: A graph with temporal extent $N_{\tau}=5$, not fitting into the 
compactified lattice and thus contributing with a negative sign on the infinite lattice.}
\label{fig_n_limit}
\end{figure}
\end{center}
We have already seen that to leading order in the strong coupling expansion 
the pressure is of ${\cal{O}}(N^0)$ in the confined phase. 
Now we will give some examples of higher orders, where cancellation of all 
$N$-dependence takes place. Let us consider at first the leading 
geometric decoration, a shift of one plaquette of the tube, 
resulting in an ${\cal{O}}(u^{4N_{\tau}+4})$ graph, 
like the one in Fig.~\ref{graphs} (left). In place of the shifted plaquette, 
we can add $v_1$, $v_2$ and $w_{ad}$ plaquettes weighted with their 
respective dimensionality. We can also put a completely new cube at this place, 
coming with a factor $-2N^2u^6$. A graph of this type is shown in 
Fig.~\ref{graphs} (middle). All possibilities summed up, we obtain the contribution
\begin{eqnarray}
P=\frac{3}{L_{\tau}a^3}u^{4N_{\tau}}\left[c_N^{N_{\tau}}+b_N^{N_{\tau}}\right]\Big[1+12N_\tau\big(1+d_{v_1}v_1+d_{v_2}v_2+d_{ad}w-2d_f^2u^{2}\big)u^4+{\cal{O}}\left(u^6\right)\Big].
\end{eqnarray}
The factor of $12N_\tau$ emerges because we can shift each of the $4N_\tau$ plaquettes in three directions. In round brackets we recognize another term $c_N$, which again cancels in the limit $N\rightarrow\infty$, and we obtain
\begin{eqnarray}
P=\frac{6}{L_{\tau}a^3}u^{4N_{\tau}}\Big[1+12N_\tau u^4+{\cal{O}}\left(u^6\right)\Big],
\end{eqnarray}
where every $N$-dependence has disappeared. 

Consider next the graphs of Fig.~\ref{fig_n_limit}, defined on the $N_{\tau}$ lattice or the infinite lattice. They are of order ${\cal{O}}(u^6)$ in the correction to the basic polymer and their contributions are
\begin{eqnarray}
\mbox{left:}\qquad\Phi_1&=&\frac{6}{L_{\tau}a^3}u^{4N_{\tau}}\bigg(-2N_{\tau}d_f^2u^6\bigg),\nonumber\\
\mbox{middle:}\qquad\Phi_2&=&\frac{6}{L_{\tau}a^3}u^{4N_{\tau}}\bigg(N_{\tau}d_f^2u^{-2}w^4\bigg),\nonumber\\
&+&\frac{6}{L_{\tau}a^3}u^{4N_{\tau}}\bigg(N_{\tau}d_f^2u^{-2}v_1^4\bigg),\nonumber\\
&+&\frac{6}{L_{\tau}a^3}u^{4N_{\tau}}\bigg(N_{\tau}d_f^2u^{-2}v_2^4\bigg),\nonumber\\
\mbox{right:}\qquad\Phi_3&=&\frac{6}{L_{\tau}a^3}u^{4N_{\tau}}\bigg(-N_{\tau}d_f^2u^6\bigg).
\end{eqnarray}
Noting that in the limit $N\rightarrow\infty$ we have \begin{eqnarray}
w=v_1=v_2=u^2,
\end{eqnarray}
we see that
\begin{eqnarray}
\Phi=\Phi_1+\Phi_1+\Phi_3=0.
\end{eqnarray}
The contributions of the left and the right graphs cancel those of the graph in the middle and the net contribution of these graphs vanishes. We remark that by leaving the limit, these graphs contribute with $1/N$ corrections.

These cancellations occur in similar ways up to ${\cal{O}}\left(u^8\right)$ 
to which we have calculated, but in higher orders they become more subtle. 
The overall picture we find is the following: fix some certain inner plaquette in a graph. 
The representation of this plaquette be of dimension $\sim N^q$. Now draw all allowed graphs, possibly consisting of several polymers, where the product of all plaquettes at these original plaquettes places is also $\sim N^q$. If we have calculated the correct contribution and taken the combinatorial factor $a(C)$ properly into account, the $N$-dependence cancels out.
To conclude, judging from our truncated series expansions, we consider the pressure to be of ${\cal{O}}\left(N^0\right)$ in the confined phase.

\section{Pressure of lattice QCD with heavy quarks}

After having calculated the leading orders of the pressure for pure lattice 
Yang-Mills theories, we want to introduce fermions in our calculations. 
The easiest way to do this is to employ the hopping parameter expansion. This is an expansion in $\kappa=(2am+8)^{-1}$ and thus a good approximation for heavy quarks.

\subsection{Hopping parameter expansion}

As the fermionic action is bilinear in the quark fields it is possible to integrate them out exactly, which we will do for Wilson fermions. Performing the Grassmann integration \cite{mm}, we find for an arbitrary flavour
\begin{eqnarray}
-S_q^f=-\ln\det(Q_f)=-\mathrm{tr}\ln\,\left(Q_f\right),
\end{eqnarray}
where the quark matrix is given by
\begin{eqnarray}
Q_f=1-\kappa_fM,
\end{eqnarray}
and the hopping matrix $M$ by
\begin{eqnarray}
M_{ab,\alpha\beta,yx}=\sum_{\mu}\delta_{y,x+\hat{\mu}}(1+\gamma_{\mu})_{\alpha\beta}U_{ab,x\mu}.
\end{eqnarray}
This leads to the hopping parameter expansion by expanding the logarithm
\begin{eqnarray}
-S_q^f=-\mathrm{tr}\,\ln\,(1-\kappa_fM)=\sum_{l}\,\frac{\kappa_f^l}{l}\,\mathrm{tr}\,M^l.
\end{eqnarray}
The delta functions in the hopping matrix $M$ force the sum to extend solely over closed loops on the lattice. 

For sufficiently strong coupling, we can omit the gluonic contribution to the partition function and write
\begin{eqnarray}
\label{hop}
Z(\kappa_f)=\int\left[dU\right] \exp\left[\sum_f\sum_{l=1}^{\infty}\frac{\kappa_f^l}{l}\mathrm{tr}\,M_f[U]^l\right].
\end{eqnarray}
In principle the sum extends over all closed loops $l$ on the lattice, which for 
finite lattice extents can have non-trivial winding numbers. In the finite temperature case, 
these are e.g.~the Polyakov loops, 
\begin{eqnarray}
L_{\vec{x}}=\mathrm{tr}\,W_{\vec{x}}=\mathrm{tr}\,\prod_{\tau=1}^{N_{\tau}}U_0(\vec{x},\tau).
\end{eqnarray}
Due to the antiperiodic boundary conditions of fermions, 
this gives an additional factor $(-1)^n$ to Eq.~(\ref{hop}), 
which depends on the winding number $n$.

One might wonder that for temporal lattice sizes $N_{\tau}\geq4$ the leading 
contributions to the effective action are those coming from the plaquette $\sim\kappa^4$. 
However, such contributions cancel in leading orders of the pressure when
subtracting the zero temperature part. The leading terms are therefore 
those from the smallest graphs with a non-trivial winding number, the Polyakov loops.
These terms are $\sim\kappa^{N_{\tau}}$ in the effective action, or 
$\sim(\kappa\mathrm{e}^{\pm a\mu})^{N_{\tau}}$ with finite chemical potential and 
the sign depending on the orientation of the Polyakov loop. 
In the confined phase the Polyakov loop cannot be screened, so the leading term in the pressure will consist of at least two (oppositely oriented) Polyakov loops, giving a factor of $\sim\kappa^{2N_{\tau}}$. This corresponds to a mesonic contribution since we used two static quark lines, one for a quark and one for an antiquark. To get a full hadron resonance gas picture, we also need baryons, which consist of three quarks or antiquarks. Hence we also have to take into account generalized Polyakov loops that wind around the temporal direction two and three times before being traced. If we do so, the effective action leading to the lowest (physical) mass hadrons reads
\begin{eqnarray}
-S^f_{eff}&=&-2(2\kappa_f)^{N_{\tau}}\sum_{\vec{x}}\Big(\mathrm{e}^{a\mu N_{\tau}}\mathrm{tr}W_{\vec{x}}+\mathrm{e}^{-a\mu N_{\tau}}\mathrm{tr}W_{\vec{x}}^{\dagger}\Big)\nonumber\\
&+&(2\kappa_f)^{2N_{\tau}}\sum_{\vec{x}}\Big(\mathrm{e}^{2a\mu N_{\tau}}\mathrm{tr}W_{\vec{x}}^2+\mathrm{e}^{-2a\mu N_{\tau}}\mathrm{tr}(W_{\vec{x}}^2)^{\dagger}\Big)\nonumber\\
&-&\frac{2}{3}(2\kappa_f)^{3N_{\tau}}\sum_{\vec{x}}\Big(\mathrm{e}^{3a\mu N_{\tau}}\mathrm{tr}W_{\vec{x}}^3+\mathrm{e}^{-3a\mu N_{\tau}}\mathrm{tr}(W_{\vec{x}}^3)^{\dagger}\Big).
\end{eqnarray}
The effective partition function is given by
\begin{eqnarray}
Z=\int\left[dU\right] \exp\Big[-S_{eff}(U)\Big]=\int\left[dU\right] \exp\Big[-\sum_fS^f_{eff}(U)\Big].
\end{eqnarray}
Expanding all terms up to ${\cal{O}}(\kappa^{3N_{\tau}})$ and doing the group integrals we get the following result for the pressure for two flavours up and down
\begin{eqnarray}
P&=&\frac{1}{L_{\tau}a^3}\bigg\lbrace4(2\kappa_u)^{2N_{\tau}}+8(2\kappa_u2\kappa_d)^{N_{\tau}}+4(2\kappa_d)^{2N_{\tau}}\bigg\rbrace\nonumber\\
&+&\frac{1}{L_{\tau}a^3}\bigg\lbrace4(2\kappa_u)^{3N_{\tau}}+6\big[(2\kappa_u)^22\kappa_d\big]^{N_{\tau}}\nonumber\\
&+&6\big[2\kappa_u(2\kappa_d)^2\big]^{N_{\tau}}+4(2\kappa_d)^{3N_{\tau}}\bigg\rbrace\Big(\mathrm{e}^{3a\mu N_{\tau}}+\mathrm{e}^{-3a\mu N_{\tau}}\Big)
\end{eqnarray}
If we now use the expressions for the hadron masses given in the appendix, we are able to rewrite this as
\begin{eqnarray}
P&=&\frac{1}{L_ta^3}\left\lbrace\sum_{0^-}\mathrm{e}^{-m\left(0^-\right)L_t}+3\sum_{1^-}\mathrm{e}^{-m\left(1^-\right)L_t}\right\rbrace\nonumber\\
&+&\frac{1}{L_ta^3}\Bigg\lbrace4\sum_{\frac{1}{2}^+}\mathrm{e}^{-m\left(\frac{1}{2}^+\right)L_t}+8\sum_{\frac{3}{2}^+}\mathrm{e}^{-m\left(\frac{3}{2}^+\right)N_t}\Bigg\rbrace\cosh\big(\mu_BL_\tau\big),\label{eq_hrg}
\end{eqnarray}
where the sum over the pseudoscalar mesons includes the pions as well as the $\eta^0$ and the vector mesons include also the $\omega^0$. This is due to the fact that we calculate with heavy quark masses and there is no reason for the pions to be lighter. We are far away from the chiral limit. The prefactors for the baryons include the spin degeneracy as well as a factor $2$ counting the corresponding antibaryons. Quark chemical potential $\mu$ has been replaced by baryon chemical potential $\mu_B=3\mu$. 

As in the case of pure Yang-Mills theories, we see that the results expected from the hadron resonance gas have been reproduced by our strong coupling and hopping parameter expansion. Similar formulas hold for a larger number of flavours. We hence conclude that at least in this parameter regime the hadron resonance gas model has been fully validated.

\section{Discussion}

In this note we have calculated the pressure of lattice Yang-Mills theories and QCD with heavy quarks in combined strong coupling and hopping parameter expansions in leading orders. Chemical potential has been introduced as an additional parameter. The leading order results are valid for all $N_\tau$ in the confined phase for small enough $\beta$. They can be used to fix the lower integration constant $P(\beta\approx0)$ in the integral method. For larger $\beta$ one should use the corresponding series expansion. In pure Yang-Mills theories we were able to calculate these expansions up to eight orders in the correction to the leading order term. In this case we could validate the hadron resonance gase model and indicate on the finiteness of the pressure in the large-$N$ limit.

In the case of QCD with heavy quarks we could also derive expressions that equal 
results of the hadron resonance gas model, albeit only in the strong coupling limit and leading orders in $\kappa$. Future work could account for $\beta$ and $\kappa$ corrections in this case. In this way one could study the effects of mass corrections, the dispersion relation as well as deviations of the ideal gas behaviour by hadron interactions.

\section*{Acknowledgement:}
We thank G.~M\"unster for numerous helpful discussions and M. Laine for a critical reading of the manuscript.
This work was supported by the BMBF project
{\em Hot Nuclear Matter from Heavy Ion Collisions and its Understanding from QCD}, 
No.~06MS254 and partly by the BMBF under project
{\em Heavy Quarks as a Bridge between Heavy Ion Collisions and QCD}.

\begin{appendix}

\section{Hadron masses from hopping parameter expansions}

In order to extract hadron masses, one has to specify suitable operators, which have non-vanishing overlap with the state, whose mass one wants to calculate. An interpolating operator for mesons $A(x)$ can be constructed from a quark and an antiquark and some $\gamma$ matrices to specify the spin structure, like for the pions and $\rho$-mesons \cite{mm}
\begin{eqnarray}
\pi_+(x)&\equiv&\bar{d}_{\alpha c}(x)\left(\gamma_5\right)_{\alpha\beta}u_{\beta c}(x),\nonumber\\
\pi_-(x)&\equiv&\bar{u}_{\alpha c}(x)\left(\gamma_5\right)_{\alpha\beta}d_{\beta c}(x),\nonumber\\
\rho_+(x)&\equiv&\bar{d}_{\alpha c}(x)\left(\gamma_k\right)_{\alpha\beta}u_{\beta c}(x),\nonumber\\
\rho_-(x)&\equiv&\bar{u}_{\alpha c}(x)\left(\gamma_k\right)_{\alpha\beta}d_{\beta c}(x).
\end{eqnarray}
The generalization to mesons with a different quark content and other vector mesons should be obvious.

For baryons or antibaryons we can do the same using three quarks or antiquarks and some $\gamma$ matrix, e.g. proton and antiproton can be constructed via
\begin{eqnarray}
p^+_{\alpha}(x)&\equiv&\varepsilon_{cde}\left(C\gamma_5\right)_{\beta\gamma}u_{\alpha c}(x)\big[u_{\beta d}(x)d_{\gamma e}(x)-d_{\beta d}(x)u_{\gamma e}(x)\big],\nonumber\\
p^-_{\delta}(y)&\equiv&\varepsilon_{fgh}\left(C\gamma_5\right)_{\varepsilon\varphi}\bar{u}_{\delta f}(y)\big[\bar{d}_{\varepsilon g}(y)\bar{u}_{\varphi h}(y)-\bar{u}_{\varepsilon g}(y)\bar{d}_{\varphi h}(y)\big],
\end{eqnarray}
with the charge conjugation Dirac matrix $C$, satisfying
\begin{eqnarray}
C\gamma_{\mu}C^{-1}=-\gamma_{\mu}^T,\qquad
-C=C^T=C^{-1}=C^{\dagger}.
\end{eqnarray}
Now we are able to calculate the masses of the lightest particles with these quantum numbers from the exponential decay of this interpolating operator in the following correlation function
\begin{eqnarray}
m(A)&=&-\mathop{\lim}_{t\rightarrow\infty}\frac{1}{t}\ln\,\langle A(0)A^{\dagger}(t)\rangle\qquad\mbox{ for mesons},\nonumber\\
m(B)&=&-\mathop{\lim}_{t\rightarrow\infty}\frac{1}{t}\ln\,\langle B(0)\bar{B}(t)\rangle\qquad\mbox{for baryons.}
\label{eq_exp_val}
\end{eqnarray}
In order to have mass eigenstates, one has to sum over all spacelike coordinates
\begin{eqnarray}
A(t)\equiv\sum_{\vec{x}}A(\vec{x},t),
\end{eqnarray}
but as we only want to calculate the leading order contribution, we need not care about this.

The masses that we obtain in this way are quite simple. They are given as
\begin{eqnarray}
\mbox{Mesons:}\qquad\quad am_{f\bar{f^\prime}}&=&-\ln2\kappa_f-\ln2\kappa_{f^\prime}\\
\mbox{Baryons:}\qquad am_{ff^\prime f^{\prime\prime}}&=&-\ln2\kappa_f-\ln2\kappa_{f^\prime}-\ln2\kappa_{f^{\prime\prime}}\\
\mbox{Antibaryons:}\qquad am_{\bar{f}\bar{f^\prime} \bar{f^{\prime\prime}}}&=&-\ln2\kappa_{\bar{f}} -\ln2\kappa_{\bar{f^\prime}}-\ln2\kappa_{\bar{f^{\prime\prime}}}.
\end{eqnarray}
We see that in leading order hopping expansion hadron masses are linear in the corresponding constituent quark masses with flavour $f$, if we define
\begin{eqnarray}
am_f=-\ln2\kappa_f\qquad\leftrightarrow\qquad\kappa_f=\frac{1}{2}\exp(-am_f),
\end{eqnarray}
relations that have been known for a long time \cite{hasen,degrand}.

\section{Strong coupling series of the Yang-Mills pressure}

The results for the pressure $P(N_\tau,u)$ of $SU(3)$ up to higher orders and small $N_\tau$ are
\begin{eqnarray}
P(1,u)a^4&=&6\,{u}^{4}+44\,{u}^{6}-30\,{u}^{7}+{\frac {2251587}{4096}}{u}^{8}-{
\frac {641763}{1024}}{u}^{9}\nonumber\\
&&+{\frac {124533967}{20480}}{u}^{10},\nonumber\\
P(2,u)a^4&=&3\,{u}^{8}+27\,{u}^{10}-135\,{u}^{11}+{\frac {1105}{4}}{u}^{12}-{
\frac {351}{4}}{u}^{13}+{\frac {18058839}{10240}}{u}^{14}\nonumber\\
&&-{\frac {
14775291}{2048}}{u}^{15}+{\frac {648558969807}{83886080}}{u}^{16},\nonumber\\
P(3,u)a^4&=&2\,{u}^{12}+54\,{u}^{14}-216\,{u}^{15}+315\,{u}^{16}-{\frac {1269}{2
}}{u}^{17}+{\frac {199458821}{30720}}{u}^{18}\nonumber\\
&&-{\frac {89855001}{5120}}
{u}^{19}+{\frac {671236701}{40960}}{u}^{20},\nonumber\\
P(4,u)a^4&=&{\frac {3}{2}}{u}^{16}+81\,{u}^{18}-297\,{u}^{19}+{\frac {1989}{4}}{
u}^{20}-{\frac {9585}{4}}{u}^{21}+{\frac {32785323}{2048}}{u}^{22}\nonumber\\
&&-{
\frac {375123069}{10240}}{u}^{23}+{\frac {2216152967}{40960}}{u}^{24},
\end{eqnarray}
and for all $N_\tau\geq5$ up to ${\cal{O}}(u^8)$ in the correction
\begin{eqnarray}
P(N_{\tau},u)a^4&=&\frac{3}{N_{\tau}}u^{4N_{\tau}}\left\lbrace c_3^{N_{\tau}}\left[1+12\,N_{\tau}{u}^{4}+42\,N_{\tau}{u}^{5}-{\frac {115343}{2048}}\,N_{\tau}{u}^{6}\right.\right.\nonumber\\
&&\left.\hspace{2cm}-{\frac {
597663}{2048}}\,N_{\tau}{u}^{7}+ \left(83\,{N_{\tau}}^
{2} +{\frac {72206061}{40960}}\,N_{\tau} \right) {u}^{8}\right]\nonumber\\
&&+b_3^{N_{\tau}}\left[1+12\,N_{\tau}{u}^{4}+30\,N_{\tau}{u}^{5}-{\frac {17191}{256}}\,N_{\tau}{u}^{6}\right.\nonumber\\
&&\left.\left.\hspace{2cm}-180\,N_{\tau}{u}^
{7}+ \left({83\,{N_{\tau}}^{2}+\frac {3819}{10}}\,N_{\tau} \right) {u}^{8}\right]\right\rbrace. 
\end{eqnarray}
Apart from the case $N_\tau=1$ all series are up to ${\cal{O}}(u^8)$. For this case the number of contributing graphs was too large to give a reliable series expansion beyond ${\cal{O}}(u^6)$.

In the limit $N\rightarrow\infty$ we find
\begin{eqnarray}
P(1,u)a^4&=&6\,{u}^{4}+44\,{u}^{6}+495\,{u}^{8}+5336\,{u}^{10},\nonumber\\
P(2,u)a^4&=&3\,{u}^{8}+94\,{u}^{12}+168\,{u}^{14}+{\frac {1983}{2}}{u}^{16},\nonumber\\
P(3,u)a^4&=&2\,{u}^{12}+72\,{u}^{16}+{\frac {404}{3}}{u}^{18}+636\,{u}^{20},\nonumber\\
P(4,u)a^4&=&{\frac {3}{2}}u^{16}+72\,u^{20}+120\,u^{22}+1121\,u^{24},\nonumber\\
P(N_\tau,u)a^4&=&\frac{6}{N_{\tau}}\,u^{4N_{\tau}}\bigg(1+12\,N_{\tau}u^4+20\,N_{\tau}u^6+ ( 83\,{N_{\tau}}^2-147\,N_{\tau}) u^8\bigg).
\end{eqnarray}
Again we have less orders for the case $N_\tau=1$. Interesting observations are the entirely positive signs and the vanishing of odd powers.

\end{appendix}

\end{document}